\documentclass[]{spie}  
\usepackage[dvips]{graphicx}
\usepackage{psfrag, amsmath}
\usepackage{makeidx}  

\title{Using Artificial Market Models to Forecast Financial Time-Series}

\author{Nachi Gupta\supit{a}, Raphael Hauser\supit{a}, and  Neil Johnson\supit{b}
\skiplinehalf
\supit{a}Oxford University Computing Laboratory, Numerical Analysis Group \\
\supit{b}Department of Physics, Oxford University
}

\authorinfo{Further author information: (Send correspondence to N.G.)  \\
E-mail: nachi@comlab.ox.ac.uk \\
N.G. would also like to thank the Clarendon Bursary for support.} 

\numberwithin{equation}{subsection}
\numberwithin{figure}{section}
\numberwithin{table}{section}
\begin{document} 
  \maketitle

\begin{abstract}
We discuss the theoretical machinery involved in predicting financial market movements using an artificial market model which has been trained on real financial data. This approach to market prediction - in particular, forecasting financial time-series by training a third-party or `black box' game on the financial data itself -- was discussed by Johnson et al. \cite{JLJHH01,LHJ02} and was based on some encouraging preliminary investigations of the dollar-yen exchange rate, various individual stocks, and stock market indices \cite{Lamp02}. However, the initial attempts lacked a clear formal methodology. Here we present a detailed methodology, using optimization techniques to build an estimate of the strategy distribution across the multi-trader population. In contrast to earlier attempts, we are able to present a systematic method for identifying `pockets of predictability' in real-world markets. We find that as each pocket closes up, the black-box system needs to be `reset' - which is equivalent to saying that the current probability estimates of the strategy allocation across the multi-trader population are no longer accurate. Instead, new probability estimates need to be obtained by iterative updating, until a new `pocket of predictability' emerges and reliable prediction can resume.
\end{abstract}

\bibliographystyle{nagpreport}


\section{Introduction}

Judging from the literature, in particular the wide range of popular finance books, the possibility of predicting future movements in financial markets ranges from significant (see, for example, the many books on chartism) to impossible\cite{Malkiel03}. Another scenario of course does exist - that financial markets may neither be predictable or unpredictable all the time, but may instead have periods where they are predictable (i.e. non-random) and periods where they are not (i.e. random).
Evidence for such `pockets of predictability' were found several years ago, by Johnson et al \cite{JLJHH01}. A similar study was reported subsequently by Sornette et al \cite{AS05}. However, a formal report of a theoretical framework for identifying such periods of predictability has not appeared in the literature to date.

The rationale behind our initial proposal to predict financial markets using artificial market models, is as follows. Financial markets produce time-series, as does any dynamical system which is evolving in time, such as the ambient air-temperature, or the electrical signals generated by heart-rhythms. In principle, one could use any time-series analysis technique to build up a picture of these statistical fluctuations and variations - and then use this technique to attempt some form of prediction, either on the long or short time-scale. One example would be to use a multivariate analysis of the prices themselves in order to build up an estimate of the parameters in the multivariate expansion, and then run this model forwards. However, such a multivariate model may not bear a relation to any physical representation of the market itself.
Instead, imagine that we are able to identify an artificial market model which seems to produce the aggregate statistical behavior (i.e. the stylized facts) observed in the financial market itself. It now has the additional advantage that it {\em also} mimics the microscopic structure of the market, i.e. it contains populations of artificial traders who use strategies in order to make decisions based on available information, and will adapt their behavior according to past successes or failures. All other things being equal, we believe that such a model may be intrinsically `better' than a purely numerical multivariate one - and may even be preferable to many more sophisticated models such as certain neural network approaches, which also may not be correctly capturing a realistic representation of the microscopic details of a physical market.  
The question then arises as to whether such an artificial market model could be `trained' on the real market in order to produce reliable forecasts.

Although in principle one could attempt to train {\em any} artificial market model on a given financial time-series, each of the model parameters will need to be estimated - and if the model has too many parameters, this will become practically impossible as the model will become over-determined. For this reason, our own (and Sornette et al. \cite{AS05}) attempts have focused on training a {\em minimal} artificial market model. We had already shown that a minimal model could be built from a binary multi-agent game in which agents are allowed to not participate if they were not sufficiently confident \cite{JLJHH01}.

Here we focus on the basic Minority Game where all agents trade at every time-step, since we are interested in describing in detail the parameter estimation process as opposed to creating the best possible market prediction model. In particular, there is no reason to expect that (i) the Minority Game's pay-off structure whereby only the minority group gets rewarded, or (ii) the Minority Game's traditional feature whereby all agents have the same memory time-scale $m$, are either realistic or optimal. We provide a more complete discussion of suitable pay-off structures in artificial financial markets in previous work\cite{JJH03,JJ02}.  For the present discussion, we retain both these features since they do not affect the formalism presented - however we note that recent work by Mitman et al. \cite{MCJ05} and subsequent work by Guo \cite{Gou05} have shown that allowing agents to have multiple memory values does indeed lead to improved performance both overall and for specific individual traders.

Binary Agent Resource games, such as the Minority Game and its many extensions, have a number of real-world applications in addition to financial markets. For example, they are well-suited to studying traffic flow in the fairly common situation of drivers having to choose repeatedly between two particular routes from work to home. In these examples, and in particular the financial market setting, one would like to predict how the system will evolve in time given the current and past states. In the context of the artificial market corresponding to the Minority Game and its generalizations, this ends up being a parameter estimation problem. In particular, it comes down to estimating the composition of the heterogeneous multi-trader population, and specifically how this population of traders is distributed across the strategy space. 
Here we investigate the use of iterative numerical optimization schemes to estimate these quantities, in particular the population's composition across the strategy space - we will then use these estimates to make forecasts on the time-series we are analyzing.  Along with these forecasts, we also need to find a covariance matrix in order to determine the certainty with which we believe our forecast is correct.  Such a covariance matrix is important for a number of reasons including risk analysis.

Given the forecast and its associated covariance matrix, we will also need to decide whether to use the forecast or throw it away based on the covariance matrix (which represents the expected errors 
on the forecast).  We discuss this point, and in so doing we will see that the system can fall into `pockets of predictability' during which the system becomes predictable over some significant time-window. In the rest of this paper, we discuss these ideas and apply them to simulated and real financial market data, in order to identify pockets of predictability based on the model.

\section{Parameterizing the Artificial Market Model}

\subsection{A Binary Agent Resource Game}

In a Binary Agent Resource game, a group of $N$ agents compete for a limited resource and each of them 
takes a binary action.  Let's denote this action at time step $k$ for agent $i$ by $a_{i}(\Omega_{k-1})$, where the action is in response 
to a global information set defined by $\Omega_{k-1}$ consisting of all information up to time step $k-1$ available to all agents.  For 
each time step, there will exist a winning decision $w_{k}$ based on the action of the agents.  This winning decision $w_{k}$ will belong 
to the next global information set $\Omega_{k}$ and will be available to each agent in the future.

\subsection{The Minority Game} \label{mg}

A particularly simple (indeed, possibly over-simplified) example of a Binary Agent Resource game is the Minority Game, which was proposed in 1997 by Challet and Zhang as a very specific game which highlights the competitive effects inherent in many complex adaptive systems.  Since then, many variants of this game have been posed with slight modifications to the original. Here we focus on the original Minority Game for the purposes of our examples, though we note that the optimization formalism which we present {\em also} applies to other variants of the game.

Let us first provide the motivation for using the Minority Game to forecast financial markets.  Essentially, in financial markets, agents compete for a limited resource, which gives us the minority nature we are interested in.  For example, if the minority group is selling, the majority group will force the price up at the following time step because of the greater demand.  At the exact same time, the minority group will sell at the overvalued price and gain profit.

We start the game with a group of $N$ agents, each of which holds an assigned strategy.  Let 
a given strategy have memory size $m$ meaning it contains a binary decision for each 
of the $2^m$ possible binary strings of length $m$.  An example of a strategy with $m=2$ is given in 
Table \ref{strat_table}.

\begin{table} 
\begin{center}
\begin{tabular}{|c|c|}
\hline
Memory & Decision \\
\hline
-1, -1 & 1 \\
-1, 1 & -1 \\
1, -1 & 1 \\
1, 1 & 1 \\
\hline
\end{tabular}
\end{center}
\caption{The left column shows the $2^2 = 4$ possible memory strings and the right column shows the strategy's response to each memory string.}
\label{strat_table}
\end{table}

A given strategy will look back at the $m$ most recent bits of the winning decisions.  At time step $k$, this would be $(w_{k-m}, \ldots, 
w_{k-1})$.  The strategy will then pick the corresponding binary decision.  Using our example strategy in Table \ref{strat_table}, if 
$w_{k-2} = -1$ and $w_{k-1} = 1$, the strategy would make the decision $-1$.  Before we can explain how the strategy works, we must also 
define the time horizon, a bit string consisting of the $T$ most recent bits of the winning decisions where $T$ is significantly larger 
than $m$.  For example, at the beginning of time step $k$, the time horizon would be $(w_{k-T}, \ldots, w_{k-1})$.  

In order for the game to be interesting, we assume some agents hold more than one strategy in what we call their strategy set.  We now need to define 
how the agents should choose amongst their strategies.  Each agent scores each of their strategies over the time horizon by giving it one 
point if it would have predicted correctly and taking one away if it would have predicted incorrectly at each time step in the past.  For example, if the time horizon was $(-1, 1, -1, -1, 1)$, we would assign $+1-1+1 = +1$ points to our strategy in Table \ref{strat_table} since the first decision on $(-1,1)$ to choose $-1$ would have been correct, while the second decision would have been incorrect and the third one correct again.  In this way, we could score all of the agents' strategies, and the agent will simply pick the highest scoring strategy to play.  The winning decision at time step $k$, $w_{k}$, is the minority decision made by the agents as a whole.

For ties between the scores of an agent's strategies, the agent will simply toss a fair coin to decide.  Further, if there is a tie in the winning decision over all agents (i.e. an equal number of agents picked -1 and 1), we can again toss a fair coin to decide.  Both of these, together, inject stochasticity into the system.  As mentioned in the introduction, there are a number of variations on ways to score strategies that can be looked at.  In this paper, we stick to the basic Minority Game structure.

We are now interested in the time-series generated by the aggregate actions of the agents.  We start the series at $r_{0} = 0$, where $r$ 
stands for returns, and allow $r_{k}$ to evolve by $r_{k} = r_{k-1} + \sum_{i}{a_{i}(\Omega_{k-1})}$, where $a_{i}(\Omega_{k-1})$ denotes the response of agent $i$ at time $k$ to global information set $\Omega_{k-1}$.  For the Minority Game, this response is simply the decision made by agent $i$, and $\Omega_{k-1}$ consists only of the time horizon at time $k-1$.  Again, agent $i$ makes this decision by choosing the highest scoring strategy over the time horizon as explained above.

Further, let's define the difference series of the returns series as $z_{0} = 0$ with $z_{k} = r_{k} - r_{k-1} =  
\sum_{i}{a_{i}(\Omega_{k-1})}$.  The difference series is the series we will estimate throughout this paper.  It is trivial to find the 
returns series given the difference series.  In terms of the difference series, we can also define the winning decisions and the time 
horizon, which we provide here for completeness.  The winning decision at time step $k$, $w_{k}$ can be defined by $w_{k} = 
\mbox{-sgn}(z_{k})$, where 
$\mbox{sgn}(k)$ represents 
the sign 
function, which is $1$ for positive $k$, $-1$ for negative $k$, and $0$ otherwise.  This simply states that when $z_{k}$ is positive, the 
minority chose $-1$ and vice versa.  Note that $z_{k}$ will never be $0$ since we toss a fair coin for ties.  As before, the time horizon 
is defined by the winning decisions.  At time step $k$, this would simply be $(w_{k-T}, \ldots, w_{k-1})$.

For a more thorough introduction to the Minority Game and other variations and extensions, there are a number of available resources. \cite{EconophysicsWebsite}

\subsection{The Parameter Estimation Problem for Forecasting} 
\label{parestprob}

Generally speaking, we would expect that the types of data we might be able to forecast using the Minority Game would have also been generated by something similar to a Minority Game. If the real market in question corresponds to a mixed majority-minority game, for example, then clearly it makes sense to attempt matching the real-data to such a mixed version of the game. This point will be explored in another publication. Here we focus on the Minority Game as a specific example.

There are many parameters in the Minority Game, so we have to choose a way to parameterize the game.  In 
this paper, we fix $m$ and $T$ for all agents.  We also say each agent possesses exactly two strategies.  
Next, we remove the parameter $N$ specifying the number of agents, by allowing this to tend towards 
infinity and instead looking at the probability distribution over all possible strategy sets with two strategies of memory size $m$.  For example, if $m=1$, there are $2^{2^1} = 4$ strategies with a memory size of one (there are $2^m$ possible bit strings of length $m$ and $2$ responses to each bit string).  So there are $\binom{4}{2} = 6$ distinct pairs of strategies with memory size $m=1$.  The parameter space we would like to estimate is the probability distribution over these six possible strategy sets.  Notice that we make a number of assumptions here to decide which parameters to estimate.  Some of these assumptions can be relaxed a bit to create a larger parameter space if desired.

In this paper, we will provide a mechanism to estimate the probability distribution over a set of strategies.  We will call this probability distribution the state $x_{k}$ at time step $k$.  In the previous paragraph, we mentioned a scheme to estimate the six pairs of $m=1$ strategies.  In this case, we were assuming all six are strategy sets that were played when generating the time-series.  However, we could also choose to estimate a strategy set with more than two strategies per agent (or maybe even just one) and each of the strategies in the set can have different memory lengths if desired (note that we may like to modify the scoring scheme for mixed memory sizes).  In this case, we would assume that this set of $N$ mixed memory and mixed strategy length strategy sets were played when generating the time-series.  

Notice that this estimation problem is an inequality constrained problem with the constraints that each probability of playing a given strategy set must be greater than or equal to 0 and the probabilities must sum up to 1.

Section \ref{iopm} and Section \ref{covmatch} discuss some of the technicalities of the optimization 
problem.  Section \ref{appsim} and Section \ref{appreal} provide some examples with results.

\section{Iterative Optimization Methods to Solve Parameter Estimation Problems} \label{iopm}

We will now look at iterative (recursive) schemes to solve time-dependent parameter estimation problems.  We desire iterative schemes for a number of reasons. They provide a forecast in only one pass of the data.  This means the algorithm can be online and can quickly make forecasts as new measurements are observed.  The iterative schemes we discuss also provide us with error bounds on forecasts so we can determine when we are in a state of high predictability (or if we ever attain such a state).  This is important for financial data since risk is always an important factor.  The first method we shall discuss is the Kalman Filter.

\subsection{Kalman Filter} \label {kf}

A Kalman Filter is simply an iterative least-squares scheme that attempts to find the best estimate at every iteration for a system governed by the following model:

\begin{equation} \label{kfsm} x_{k} = \Phi_{k,k-1} x_{k-1} + u_{k}, \qquad u_{k} \sim N(0,Q_{k,k-1}) \end{equation}
\begin{equation} \label{kfmm} z_{k} = H_{k} x_{k} + v_{k}, \qquad v_{k} \sim N(0,R_{k}) \end{equation}

Here $x_{k}$ represents the true state of the underlying system.  $\Phi_{k,k-1}$ represents the matrix used to make the transition from state $x_{k-1}$ to $x_{k}$.  The variable $z_{k}$ represents the measurement (or observation).  $H_{k}$ is the matrix that takes the state into measurement space.  The variables $u_{k}$ and $v_{k}$ are both noise terms which are normally distributed with mean 0 and variances $Q_{k,k-1}$ and $R_{k}$, respectively.  

The Kalman Filter will at every iteration make a prediction for $x_k$ which we denote by $\hat{x}_{k|k-1}$.  We use the notation ${k|k-1}$ since we will only use measurements provided until time step $k-1$ to make the prediction at time $k$.  We can define the state prediction error $\tilde{x}_{k|k-1}$ as the difference between the true state and the state prediction.

\begin{equation} \label{se1}
\tilde{x}_{k|k-1} = x_{k} - \hat{x}_{k|k-1}
\end{equation}

In addition, the Kalman Filter will provide a state estimate for $x_{k}$ given all the measurements provided up to and including time step $k$.  We denote these estimates by $\hat{x}_{k|k}$.  We can similarly define the state estimate error by

\begin{equation} \label{se2}
\tilde{x}_{k|k} = x_{k} - \hat{x}_{k|k}
\end{equation}

Since we assume $u_{k}$ is normally distributed with mean $0$, we make the state prediction simply by using $\Phi_{k,k-1}$ to make the 
transition.  This is given by

\begin{equation} \label{kfsp} \hat{x}_{k|k-1} = \Phi_{k,k-1} \hat{x}_{k-1|k-1} \end{equation}

We can also calculate the associated covariance for the state prediction, which we call the covariance prediction.  This is actually just 
the expectation of the outer product of the state prediction error with itself. This is given by

\begin{equation} \label{kfcp} P_{k|k-1} = \Phi_{k,k-1} P_{k-1|k-1} \Phi_{k,k-1}' + Q_{k,k-1} \end{equation}

Notice that we use the prime notation on a matrix throughout this paper to denote the transpose of that matrix.  Now we can make a prediction on what we expect to see for our measurement, which we call the measurement prediction by

\begin{equation} \label{kfmp} \hat{z}_{k|k-1} = H_{k} \hat{x}_{k|k-1} \end{equation}

The difference between our true measurement and our measurement prediction is often times called the innovation (or measurement residual).  We will use the term innovation throughout this paper, and we calculate this by

\begin{equation} \label{kfi} \nu_{k} = z_{k} - \hat{z}_{k|k-1} \end{equation}

We can also calculate the associated covariance for the innovation, which we call the innovation covariance, by

\begin{equation} \label{kfic} S_{k} = H_{k} P_{k|k-1} H_{k}' + R_{k} \end{equation}

Next, we will calculate the Kalman Gain, which lies at the heart of the Kalman Filter.  This essentially tells us how much we prefer our new measurement over our measurement residual.  We calculate this by

\begin{equation} \label{kfkg} K_{k} = P_{k|k-1} H_{k}' S_{k}^{-1} \end{equation}

Using the Kalman Gain and the innovation, we update the state estimate.  If we look carefully at the following equation, we are essentially taking a weighted sum of our state prediction with the Kalman Gain multiplied by the innovation.  So the Kalman Gain is telling us how much to ``weight in'' information contained in the new measurement.  We calculate the updated state estimate by

\begin{equation} \label{kfsu} \hat{x}_{k|k} = \hat{x}_{k|k-1} + K_{k}  \nu_{k} \end{equation}

Last but not least, we calculate the updated covariance estimate.  This is actually just the expectation of the outer product of the 
state error estimate with itself.  Here we will 
give the most numerically stable form of this equation, as this form prevents loss of symmetry and best preserves positive definiteness

\begin{equation} \label{kfcu} P_{k|k} = (I - K_{k} H_{k}) P_{k|k-1} (I - K_{k} H_{k})' + K_{k} R_{k} K_{k}^{T} \end{equation}

The covariance matrices throughout the Kalman Filter give us a way to measure the uncertainty of our state prediction, state estimate, 
and the innovation.  Also, notice that the 
Kalman Filter is recursive, and we require an initial estimate $\hat{x}_{0|0}$ and associated covariance matrix 
$P_{0|0}$.    Here we simply provided the equations of the Kalman Filter without derivation.  For a more thorough understanding of the Kalman Filter, there are a number of available resources \cite{BLK01}.

\subsection{Constrained Iterative Optimization Methods} \label {const_op}

A Kalman Filter would certainly be the correct tool for the parameter estimation problem described in \ref{parestprob} if we were interested in an iterative solution and did not have any equality and inequality constraints.  However, note that we have the following constraints on our states at each time step that make this a constrained problem:

\begin{equation} \label{constraints}
\sum_{i} {\hat{x}_{i,k|k}} = 1 \mbox{ and } \hat{x}_{i,k|k} \geq 0, \forall i
\end{equation}

Here $\hat{x}_{i,k|k}$  is the $i$-th element of $\hat{x}_{k|k}$, which represents the single probability of using a certain strategy set at time step $k$.  Since we would like to use an iterative scheme, we must now think of a different method which acts as a Kalman Filter but allows for equality and inequality constrained optimization.  In Section \ref{eq_const}, we will introduce a method for solving equality constrained problems iteratively in a Kalman Filter like manner.  From here we will make the extension to inequality constrained problems in Section \ref{ineq_const}.

\subsection{Nonlinear Equality Constraints} \label{eq_const}

Let's add to our model given by equations \eqref{kfsm} and \eqref{kfmm} the following smooth nonlinear equality constraints

\begin{equation} \label{Lxeq0} e_{k}(x_{k}) = 0 \end{equation}

Notice that our constraints provided in equation \eqref{constraints} are actually linear.  We present the nonlinear case for further completeness here.  We now rewrite the problem we would like to solve where we use the superscript $c$ to denote constrained.  We should also rephrase the problem we would like to solve now.  We are given the last prediction and its covariance, the current measurement and its covariance, and a set of equality constraints and would like to make the current prediction and find its covariance matrix.

Let's write the problem we are solving as

\begin{equation} \label{eq1}  z_{k}^{c} = h_{k}^{c}(x_{k}) + v_{k}^{c}, \qquad v_{k}^{c} \sim N(0,R_{k}^{c}) \end{equation}

Here $z_{k}^{c}$, $h_{k}^{c}$, and $v_{k}^{c}$ are all vectors, each having three distinct parts.  The first part will represent the prediction for the current time step, the second part is the measurement, and the third part is the equality constraint.  $z_{k}^{c}$ effectively still represents the measurement, with the prediction treated as a ``pseudo-measurement" with its associated covariance.

\begin{equation} \label{eq2}
z_{k}^{c} = \left[ \begin{array}{c}
	\Phi_{k,k-1} \hat{x}_{k-1|k-1} \\ 
	z_{k} \\
	0
\end{array}\right]
\end{equation}

The matrix $h_{k}^{c}$ takes our state into the measurement space as before

\begin{equation} \label{eq3}
h_{k}^{c}(x_{k}) = \left[ \begin{array}{c} 
	x_{k} \\
	H_{k} x_{k} \\
	e_{k}(x_{k}) 
\end{array} \right]
\end{equation}

Notice that by combining equations \eqref{se1} and \eqref{se2}, we can rewrite the state error prediction as 

\begin{equation} \label {se3}
\tilde{x}_{k|k-1} = \Phi_{k,k-1} \tilde{x}_{k-1|k-1} + u_{k-1}
\end{equation}

Now we can define $v_{k}^{c}$ again as the noise term using equation \eqref{se3}.  

\begin{equation} \label{eq4}
v_{k}^{c} = \left[ \begin{array}{c}
	-\Phi_{k,k-1} \tilde{x}_{k-1|k-1} - u_{k-1} \\
	v_{k} \\
	0
\end{array} \right]
\end{equation}

And $v_{k}^{c}$ will be normally distributed with mean 0 and variance $R_{k}^{c}$.  The diagonal elements of $R_{k}^{c}$ represent the variance of each element of $v_{k}^{c}$.  We define the covariance of the state estimate error at time step $k$ as $P_{k|k}$.  Notice also that $R_{k}^{c}$ contains no off diagonal elements.

\begin{equation} \label{eq5}
R_{k}^{c} = \left[ \begin{array}{ccc}
	\Phi_{k,k-1} P_{k-1|k-1} {\Phi_{k,k-1}}' + Q_{k,k-1} & 0 & 0 \\
	0 & R_{k} & 0 \\
	0 & 0 & 0
\end{array} \right]
\end{equation} 

This method of expressing our problem can be thought of as a fusion of the state prediction and the new measurement at each iteration under the given equality constraints.  Much like when we showed the Kalman Filter, we will simply write the solution here \cite{CWCP02,WCC02}.

\begin{equation} \label{xeq}
\hat{x}_{k|k,j} = \left[ \begin{array}{cc}
	0 & I
\end{array} \right] \left[ \begin{array}{cc}
	R_{k}^{c} & H_{k,j}^{c} \\
	{H_{k,j}^{c}}' & 0
\end{array} \right]^{+} \left[ \begin{array}{c}
	z_{k}^{c} - h_{k}^{c}(\hat{x}_{k|k,j-1}^{c}) + H_{k,j}^{c} \hat{x}_{k|k,j-1}^{c} \\
	0
\end{array} \right]
\end{equation}

Notice the we use the $^{+}$ notation on a matrix throughout this paper to denote the pseudo-inverse of that matrix.  This method 
significantly differs from a Kalman Filter.  In this method we are iterating over a dummy variable $j$ within each time step 
until we fall within a predetermined convergence bound $\left| \hat{x}_{k|k,j} - \hat{x}_{k|k,j-1} \right| \leq c_k$ or hit a chosen number of maximum iterations.  We initialize our first iteration as $ \hat{x}_{k|k,0} = \hat{x}_{k-1|k-1} $ and use the final iteration as $ \hat{x}_{k|k} = \hat{x}_{k|k,J}$ where $J$ represents the final iteration.

Also, notice that we allowed the equality constraints to be nonlinear.  As a result, we define $H_{k,j}^{c} = \frac{\partial h_{k}^{c}}{\partial x_{k}}(\hat{x}_{k|k,j-1})$ which gives us a local approximation to the direction of $h_{k}^{c}$.

We actually find a stronger form for this solution \cite{CWCP02,WCC02}, where $R_{k}^{c}$ will reflect the tightening of the covariance for the state prediction based on the new estimate at each iteration of $j$.  We do not tighten the covariance matrix within these iterations here, since in our form, we can actually change the number of equality constraints between iterations of $j$.  We will find this useful in the next section.  Not tightening the covariance matrix in this way is reflected in a larger covariance matrix for the estimate as well.  This covariance matrix is calculated as

\begin{equation} \label{Peq}
P_{k|k,j} = - \left[ \begin{array}{cc}
	0 & I
\end{array} \right] \left[ \begin{array}{cc}
	R_{k}^{c} & H_{k,j}^{c} \\
	{H_{k,j}^{c}}' & 0
\end{array} \right]^{+} \left[ \begin{array}{c}
	0 \\
	I
\end{array} \right]
\end{equation}

Notice that for faster computation times, we need only calculate $P_{k|k,j}$ for the final iteration of $j$.  Further, if our equality constraints are in fact independent of $j$, we can calculate $H_{k,j}^{c}$ only once for each $k$.  This would also imply the pseudo-inverse in equation \eqref{xeq} can be calculated only once for each $k$.

This method, while very different from the Kalman Filter presented earlier, provides us with an estimate $\hat{x}_{k|k}$ and a covariance matrix for the estimate $P_{k|k}$ at each time step similar to the Kalman Filter.  However, this method allowed us to incorporate equality constraints.

\subsection{Nonlinear Inequality Constraints} \label{ineq_const}

We will now extend the equality constrained problem to an inequality constrained problem.  To our system given by equations \eqref{kfsm}, \eqref{kfmm}, and \eqref{Lxeq0}, we will also add the smooth inequality constraints given by

\begin{equation} l_{k}(x_{k}) \geq 0. \end{equation}

Our method will be to keep a subset of the inequality constraints active at any time.  An active constraint is simply a constraint that we treat as an equality constraint.  An inactive constraint we will relax (ignore) when solving our optimization problem.  After, solving the problem, we then check if our solution lies in the space given by the inequality constraints.  If it doesn't we start from the solution in our previous iteration and move in the direction of the new solution until we hit a set of constraints.  For the next iteration, this set of constraints will be the new active constraints.

We formulate the problem in the same way as before keeping equations \eqref{eq1}, \eqref{eq2}, \eqref{eq4}, and \eqref{eq5} the same to set up the problem.  However, we replace equation \eqref{eq3} by

\begin{equation}
h_{k}^{c}(x_{k}) = \left[ \begin{array}{c} 
	x_{k} \\
	H_{k} x_{k} \\
	e_{k}(x_{k}) \\
	l_{k,j}^{a}(x_{k})
\end{array} \right]
\end{equation}

$l_{k,j}^{a}$ represents the set of active inequality constraints.  Notice that while we keep equations \eqref{eq2}, \eqref{eq4}, and \eqref{eq5} the same, these will need to be padded by additional zeros appropriately to match the size of $l_{k,j}^{a}$.  Now we solve the equality constrained problem consisting of the equality constraints and the active inequality constraints (which we treat as equality constraints) using equations \eqref{xeq} and \eqref{Peq}.  However, let's call the solution from equation \eqref{xeq} $\hat{x}_{k|k,j}^{*}$ since we have not checked if this solution lies in the inequality constrained space yet.  In order to check this, we find the vector that we moved along to reach $\hat{x}_{k|k,j}^{*}$.  This is simply 

\begin{equation} d = \hat{x}_{k|k,j}^{*} - \hat{x}_{k|k,j-1} \end{equation}

We now iterate through each of our inequality constraints to check if they are satisfied.  If they are all satisfied, we choose $t_{\max}=1$, and if they are not, we choose the largest value of $t_{\max}$ such that $\hat{x}_{k|k,j-1} + t_{\max} d$ lies in the inequality constrained space.  We choose our estimate to be

\begin{equation} \hat{x}_{k|k,j} = \hat{x}_{k|k,j-1} + t_{\max} d \end{equation}

We also would like to remember the inequality constraints which are being touched in this new solution.  These constraints will now become active for the next iteration and lie in $l_{k,j+1}^{a}$.  Note that $l_{k,0}^{a} = l_{k-1,J}^{a}$, where $J$ represents the final iteration of a given time step.

Note also that we do not perturb the error covariance matrix from equation \eqref{Peq} in any way.  Under the assumption that our model is a well-matched model for the data, enforcing inequality constraints (as dictated by the model) should only make our estimate better.  Having a slightly larger covariance matrix is better than having an overly optimistic one based on a bad choice for the perturbation \cite{SS03}.  Perturbing this covariance matrix correctly may be investigated in the future.

\section{Covariance Matching Techniques} \label{covmatch}

In many applications of Kalman Filtering, the process noise $Q_{k,k-1}$ and measurement noise $R_{k}$ are known.  However, in our application we are not provided with this information a priori so we would like to estimate them.  These can often times be difficult to approximate especially when there is a known model mismatch.  We will present one possible method to approximate these \cite{Mayb79}.  We choose to match the process noise and measurement noise to the past innovation (residual) process.

\subsection{Determining the Process Noise and Measurement Noise}

In addition to estimating $Q_{k,k-1}$ and $R_{k}$, we will also estimate the innovation covariance $S_{k}$.  We can actually determine the innovation covariance from equation \eqref{kfic}, but estimating it using covariance matching to the past innovation process can provide us with a more accurate innovation covariance.

We estimate $S_{k}$ by taking a window of size $N_{k}$ (which is picked in advance for statistical smoothing) and time-averaging the innovation covariance based on the innovation process.  This is simply the average of all the outer products of the innovations over this window.

\begin{equation} \label{Sk}
\hat{S}_{k}^{*} = \frac{1}{N_{k}-1} \sum_{j=k-N_{k}}^{k-1}{\nu_{j} {\nu_{j}}'}
\end{equation}

Next, let's estimate $R_{k}$.  This is done similarly.  If we refer back to equation \eqref{kfic}, we can simply calculate this by

\begin{equation} \label{Rk}
\hat{R}_{k}^{*} = \frac{1}{N_{k}-1} \sum_{j=k-N_{k}}^{k-1}{\nu_{j} {\nu_{j}}' - H_{j} P_{j|j-1} H_{j}'}
\end{equation}

We can now use our choice of $R_{k}$ along with our innovation covariance $S_{k}$ to estimate $Q_{k,k-1}$.  Combining equations \eqref{kfcp} and \eqref{kfic} we have

\begin{equation}
S_{k} = H_{k} (\Phi_{k,k-1} P_{k-1|k-1} {\Phi_{k,k-1}}' + Q_{k,k-1}) {H_{k}} + R_{k}
\end{equation}

Bringing all $Q_{k,k-1}$ terms to one side leaves us with

\begin{equation}
H_{k} Q_{k,k-1} {H_{k}}' = S_{k} - H_{k} \Phi_{k} P_{k-1|k-1} {\Phi_{k}}' {H_{k}}' - R_{k}
\end{equation}

And solving for $Q_{k,k-1}$ gives us

\begin{equation} \label{Qk}
\hat{Q}_{k,k-1}^{*} = {\left({H_{k}}' H_{k}\right)}^{+} {H_{k}}'  \left( S_{k} - H_{k} \Phi_{k} P_{k-1|k-1} {\Phi_{k}}' {H_{k}}' - R_{k} \right) H_{k} {\left({H_{k}}' H_{k}\right)}^{+}
\end{equation}

Note that it may be desirable to keep $\hat{Q}_{k,k-1}^{*}$ diagonal if we do not believe the process noise has any cross-correlation.  It is rare that you would expect a cross-correlation in the process noise.  In addition, keeping the process noise diagonal has the effect of making our covariance matrix ``more positive definite.''  This can be done simply by setting the off diagonal terms of $\hat{Q}_{k,k-1}^{*}$ equal to $0$.

It is also important to keep in mind that we are estimating covariance matrices here which must be symmetric and positive semidefinite (note that the diagonal elements should always be greater than or equal to zero as these are variances).

\subsection{Upper and Lower Bounds for Covariance Matrices}

We might also like to denote a minimum and maximum number we are willing to accept for each element of our covariance matrices. The motivation for maintaining a minimum is that we may not want to become overly optimistic.  If the covariances drop to zero, we will assume the random variable has perfect knowledge.  The reason for maintaining a maximum is in case we believe the covariance actually is upper bounded.  Let us denote these matrices by $S_{k}^{\min}$, $R_{k}^{\min}$, $Q_{k}^{\min}$,  $S_{k}^{\max}$, $R_
{k}^{\max}$, and $Q_{k}^{\max}$.  We apply these by

\begin{equation} \label{sk2}
\hat{S}_{k} = \frac{1}{N_{k}-1} \sum_{j=k-N_{k}}^{k-1} {\min\left( S_{j}^{\max}, \max\left(S_{j}^{\min}, \nu_{j} {\nu_{j}}'\right) \right)},
\end{equation}

\begin{equation} \label{rk2}
\hat{R}_{k} = \frac{1}{N_{k}-1} \sum_{j=k-N_{k}}^{k-1}{\min\left(R_{j}^{\max}, \max\left(R_{j}^{\min}, \nu_{j} {\nu_{j}}' - H_{j} P_{j|j-1} H_{j}'\right) \right)},
\end{equation}

and, using equation \eqref{Qk},

\begin{equation}
\hat{Q}_{k} = \min\left(Q_{k}^{\max}, \max\left(Q_{k}^{\min}, \hat{Q}_{k}^{*}\right) \right)
\end{equation}

Again, keep in mind that the diagonal elements of $S_{k}^{\min}$, $R_{k}^{\min}$, and $Q_{k}^{\min}$ must all be greater than or equal to zero.  This is a very simple way of lower and upper bounding these matrices.  There will certainly be a number of ways we could approach this problem some of which might be much better at preserving the original information.  Our hope for the application mentioned in this paper is that the bounds are rarely touched if ever.

\section{Application of Inequality Constrained Iterative Optimization Methods to a Simulated Minority Game} \label{appsim}

In this section, we will apply the discussed methods to a simulation of the Minority Game.  In the next section, we will apply these methods to real financial data.

\subsection{Generating Simulation Data} \label{appsim-gsd}

We choose parameters $m=1$ for the memory size and allow two strategies per agent resulting in six overall combinations as described in Section \ref{parestprob}.  Also, we choose the time horizon size to be $T=50$.  We randomly choose a bit string of length $50$ to serve as the initial time horizon, and we also randomly choose a probability distribution over the six possible strategy sets.  We run the simulation over 150 time steps.  This results in a returns series $r_{k}$ from which we can extract the difference series $z_{k}$.

\subsection{Predicting the Simulated Market Data} \label{psd}

\subsubsection{Forming the Estimation Problem}
To track the difference series $z_{k}$, we set the problem up similar to how it was generated with $m=1$ giving us six probabilities, and we choose the time horizon size to be $T=50$.  

We also make the assumption that the estimate for the probability distribution at time $k$ will also be the prediction at time $k+1$ since we hope that our estimate at time $k$ will be well matched to the data locally.  For the simulated case, the probability distribution is actually fixed over all $k$.  This boils down to choosing the identity matrix as the transition matrix ($\Phi_{k,k-1}$ for all $k$ in the notation of Section \ref{kf}).

We create the time horizon at each time step by looking back $T$ time steps and checking where the minority would have lied at each time step as described near the end of Section \ref{mg} using the difference series $z_{k}$.  If an element of $z_{k}$ is 0 (implying there was no minority), we simply skip this point in our time horizon.

Finally, we score each strategy in each of our strategy sets over the time horizon similar to what we described in Section \ref{mg}.  This will result in a set of winning strategies.  We determine the set of predictions of each of the winning strategies and this forms the measurement matrix $H_{k}$ (it is actually a vector of $\pm 1$ since the measurements are scalars).  Multiplying this by the state gives us the prediction based on the probability of being in a certain strategy set and what the set would pick as its forecast.

Notice that in most tracking applications, the transition matrix $\Phi_{k,k-1}$ drives the system evolution through time and the measurement matrix $H_{k}$ describes a fixed coordinate transform.  Here $H_{k}$ changes significantly based on the state estimate $\hat{x}_{k-1|k-1}$ of the system.   In fact, the process noise $Q_{k,k-1}$ and measurement noise $R_{k}$, found by covariance matching techniques in our case, combined with $H_{k}$ are really what drive the system evolution through time.

\subsubsection{Choosing the minimum and maximum acceptable covariances}
For the minimum and maximum acceptable covariances used in the covariance matching scheme of Section \ref{covmatch}, we choose 

\begin{equation} \label{SR_cov}
S_{k}^{\min} = R_{k}^{\min} = 0 \mbox{ and } S_{k}^{\max} = R_{k}^{\max} = 1
\end{equation}

Note that both the measurements and the innovations must lie in $[-1,1]$.  This is true because in the extreme situations, all the 
strategies can choose either $-1$ or $1$.  It is also known that the variance of the distribution with half of its weight at $a$ and the 
other half at $b$ is given by $\frac{(b-a)^{2}}{4}$.  Applying this formula gives us the maximum acceptable variances of $1$ in equation 
\eqref{SR_cov}.  We use a similar idea to choose the minimum and maximum acceptable process noise.

\begin{equation}
Q_{k}^{\min} = \left[ \begin{array}{cccc}
0 & 0 & \cdots & 0 \\
0 & 0 & \cdots & 0 \\
\vdots & \vdots & \ddots & \vdots \\
0 & 0 & \cdots & 0
\end{array} \right]
\end{equation}

and 

\begin{equation} \label{Qkmax}
Q_{k}^{\max} = \left[ \begin{array}{cccc}
.25 & 0 & \cdots & 0 \\
0 & .25 & \cdots & 0 \\
\vdots & \vdots & \ddots & \vdots \\
0 & 0 & \cdots & .25
\end{array} \right]
\end{equation}

We choose $.25$ for the diagonal terms of $Q_{k}^{max}$ by our previous logic since each element of the probability distribution must lie 
in $[0,1]$.  We force 
the diagonal elements to be $0$ in order to keep our covariances more positive definite.

In addition, we state the following definition

\begin{equation} \label{cor1}
\mbox{cov}(x,y) =  \mbox{cor}(x,y) \sigma_{x}\sigma_{y}
\end{equation}

Noticing that the $\mbox{cor}(x,y)$ takes its most extreme values at $\pm 1$ and $\sigma_{x}$ and $\sigma_{y}$ both take their largest values at $\frac{b-a}{2}$, we can state

\begin{equation} \label{correl}
\left|\mbox{cov}(x,y)\right| \leq {\left(\frac{b-a}{2}\right)}^{2}
\end{equation}

We may also be interested in bounding our state prediction error and state estimate error covariances since we know 
we are estimating a probability distribution.  Using equation \eqref{correl} for the off diagonal terms, we can bound these by

\begin{equation} \label{Pmin}
P_{k|k-1}^{\min} = P_{k|k}^{\min} = \left[ \begin{array}{cccc}
0 & -.25 & \cdots & -.25 \\
-.25 & 0 & \cdots & -.25 \\
\vdots & \vdots & \ddots & \vdots \\
-.25 & -.25 & \cdots & 0
\end{array} \right]
\end{equation}

and 

\begin{equation} \label{Pmax}
P_{k|k-1}^{\max} = P_{k|k}^{\max} = \left[ \begin{array}{cccc}
.25 & .25 & \cdots & .25 \\
.25 & .25 & \cdots & .25 \\
\vdots & \vdots & \ddots & \vdots \\
.25 & .25 & \cdots & .25
\end{array} \right]
\end{equation}

We did not provide a very rigorous explanation for our choice of covariance bounds here.  However, in our example, these 
are the choices we made for the reasons provided above.

For the covariance matching, we also choose 
$N_{k}$ in equations \eqref{sk2} and \eqref{rk2} to be equal to $T=50$.  Notice that while we have less than $50$ innovations, we choose $N_{k}$ to be the number of innovations we have.  When we have $0$ innovations (at the initial point), we choose $R_{k} = 0$ so we can heavily trust the first measurement to strengthen our initialization.

\subsubsection{Initial Parameters} \label{ip}
We choose our initial state $\hat{x}_{0|0} = \frac{1}{s} 1_{s}$, where $s$ is the number of strategy sets (in our case $6$) and $1_{s}$ is a column vector of size $s$ full of $1$'s.  This is essentially starting with a uniform distribution.  We also choose our initial covariance $P_{0|0} = .25 I_{s \times s}$, where $I_{s \times s}$ represents the $s \times s$ identity matrix.  We choose $.25$ again for the same reason as before.  Note that we will actually start the optimization problem at time step $T+1$ since we use the first $T$ data points to initialize the time horizon.  We also assign no process noise initially and zero measurement noise on the first measurement.

Using the methods described in Section \ref{ineq_const} and Section \ref{covmatch}, we can now make predictions on this system.  After making predictions, we need a system by which to decide which predictions to accept with greater certainty.  We discuss this in the next section.

\subsection{Effective Forecasting} \label{ef}

The last question we would like to address here is: when is the forecast produced by this method good and how good?  We could base this on the innovation covariance $S_k$ which is an estimate of the errors of the innovation (residual) process.  Note that we can either use equation \eqref{kfic} along with equation \eqref{kfcp} or we can use equation \eqref{Sk} to calculate $S_{k}$.    The residual-based estimate given by equation \eqref{Sk} will generally provide a smoother function through $k$ which might be desirable to find pockets of predictability (where we can predict well for a while).

There are various ways of using $S_{k}$ to decide when we would like to make a prediction.  We look at a very simple method, where we simply take a threshold value $t_{k}$ such that we choose $k$ in our set of prediction times if $S_{k} \leq t_{k}$.

\subsection{Results of simulation} \label{simresult}

\begin{figure}[h!] 
\begin{center}
\psfrag{Simulated Data}{Simulated Data}
\psfrag{Innovations}{Innovations}
\psfrag{Prediction Times}{Prediction Times}
\includegraphics[height=4in,width=\textwidth]{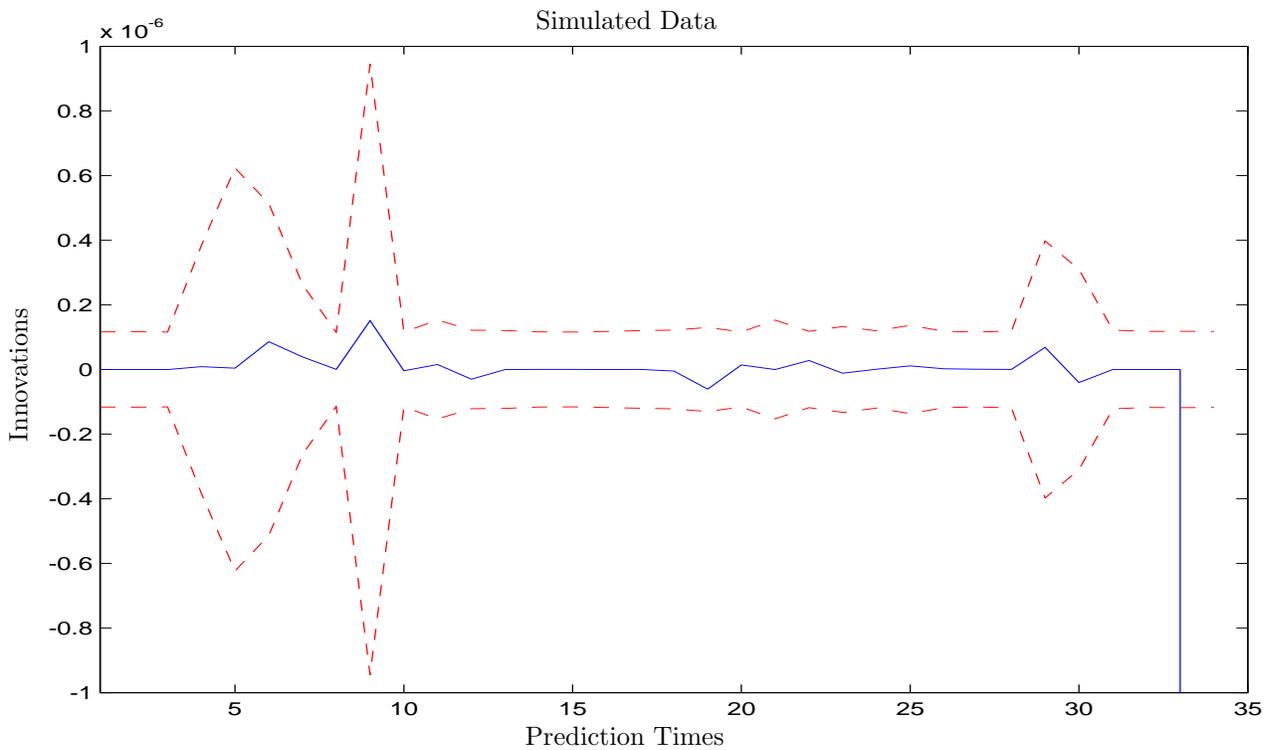}
\end{center}
\caption{In the above plot, the solid line represents the innovation (measurement residual) process and the dashed line represents one standard deviation about zero based on the predicted innovation variance.  We select ``Prediction Times" in this plot as times when the data is in a predictable state based on the innovation covariance.  These times need not be consecutive in the original data although often times are.}
\label{fig1}
\end{figure}

We show the results from the simulated data in Figure \ref{fig1}.  We choose the threshold value $t_{k}$ in this plot to be $10^{-3}$ for all $k$.  Notice that in our case, $t_k$ is a scalar since the measurements are scalars.  As we can see in the plot, we are able to make good forecasts at over $30$ points (where our innovations lie within the innovation standard deviation).  Notice that we only attempt to make forecasts at the last $100$ points of the $150$ generated data points (we use the first $50$ points to generate the initial time horizon as mentioned earlier).  However, we choose only to make a prediction at $34$ of the data points.  It happens to be that these $34$ data points are the first $34$ that we attempt to forecast.  We have $2$ bad predictions at the end of the plot.  After the bad predictions, we never recover to a good prediction since the covariance matching scheme drives the estimated process noise $\hat{Q}_{k,k-1}$, estimated measurement noise $\hat{R}_{k}$, and estimated innovation covariance $\hat{S}_{k}$ up due to the large spike in the single residual which affects the statistical smoothing for $50$ time steps.  Since the covariance of the state remains tight and the covariance of the measurements is relatively large, new measurements aren't trusted and given much weight for creating forecasts.

At the same time, because of the transient by the statistical smoothing, we continue to make forecasts immediately after the first false prediction.  We might choose to incorporate a scheme to not make predictions for some length of time immediately after a false prediction to allow $\hat{Q}_{k,k-1}$, $\hat{R}_{k}$, and $\hat{S}_{k}$ to respond to the shock caused by the false prediction.  Further, we might like to significantly increase the process noise after a false prediction to effectively cause new measurements to have a stronger weight in forming estimates.

\section{Application of Inequality Constrained Iterative Optimization Methods to Real Foreign Exchange Data} \label{appreal}

We now apply the ideas in this paper to real financial data.  We choose hourly USD/YEN foreign exchange rate data from 1993 to 1994 provided by Dr. J. James of Bank One in London.

\subsection{Setting up the Problem}

\subsubsection{Scaling the difference series}
We first find the difference series from the returns series as before.  For the time being, let's call this $z_{k}^{*} = r_{k} - r_{k-1}$.  Since our algorithm only makes forecasts in $[-1,1]$, we might like to scale our inputs to this domain as best as possible.  Assuming that we have measurements a priori up to time step $K$, we estimate the scaling based on these measurements.  Let's denote the set of all measurements up to time $K$ by $z_{K}^{*}$ where $z_{K}^{*}$ in our case is a vector of scalar measurements.  We choose the following method:

Let's denote the minimum and maximum elements of  vector $V$ by the functions $\min(V)$ and $\max(V)$, respectively.  And let's denote the minimum and maximum elements of $z_{K}^{*}$ by $z_{K}^{\min}$ and $z_{K}^{\max}$, respectively.  We first proportionally scale the spacing between elements of $z_{k}^{*}$ so the difference between the minimum element and the maximum element is $2$ (the size of $[-1,1]$).  We denote this by $z_{k}^{**}$

\begin{equation} \label{scale1}
z_{k}^{**} = \frac{2z_{k}^{*}}{z_{K}^{\max} - z_{K}^{\min}}
\end{equation}

Next, we scale the elements so the minimum element is at $-1$.  This will automatically place the maximum element at $+1$.

\begin{equation} \label{scale2}
z_{k} = z_{k}^{**} - (\min(z_{k}^{**})+1)
\end{equation}

We do the calculation for $z_{K}^{\min}$ and $z_{K}^{\max}$ once with all of the a priori information we have.  Then we can use equations \eqref{scale1} and \eqref{scale2} to scale for any time step $k$.  Our hope is that based on the a priori information, our choice of $z_{K}^{\min}$ and $z_{K}^{\max}$ will reflect the true spacing.  If we know true values for these, we can use them instead.  Notice also that we can still find the returns series $r_{k}$ from this definition for the measurements simply by inverting the process.

There will certainly be a number of different ways to do this scaling as well.

\subsubsection{Forming the Estimation Problem}
For the rest of this estimation problem, we actually do the setup exactly the same as in Section \ref{psd} and we follow the effective forecasting scheme exactly as in Section \ref{ef} choosing threshold value $t_{k}$ to be $10^{-3}$ again for all $k$.

\subsection{Results on the Real Data}
\begin{figure}[h!]
\begin{center}
\psfrag{Hourly USD/YEN FX-rate from 1993 to 1994}{Hourly USD/YEN FX-rate from 1993 to 1994}
\psfrag{Innovations}{Innovations}
\psfrag{Prediction Times}{Prediction Times}
\includegraphics[height=4in,width=\textwidth]{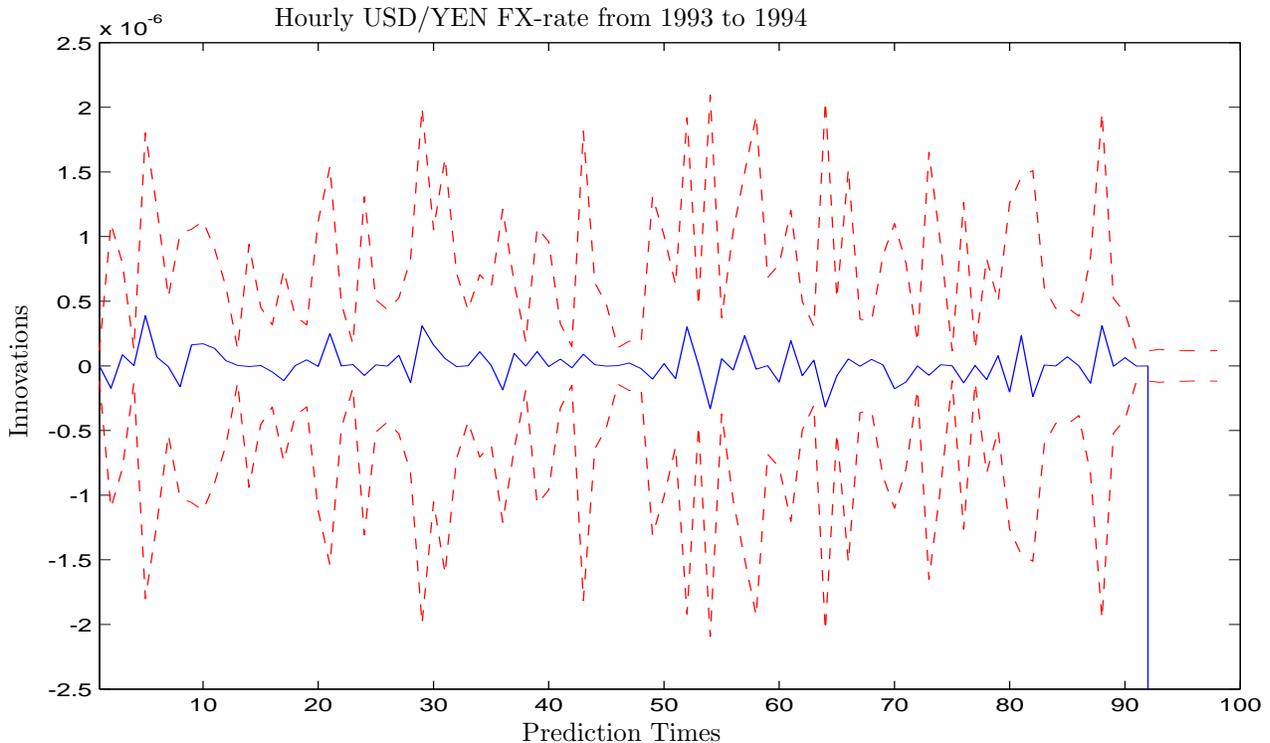}
\end{center}
\caption{In the above plot, the solid line represents the innovation (measurement residual) process and the dashed line represents one standard deviation about zero based on the predicted innovation variance.  We select ``Prediction Times" in this plot as times when the data is in a predictable state based on the innovation covariance.  These times need not be consecutive in the original data although often times are.}
\label{fig2}
\end{figure}

We show the results from the real data in Figure \ref{fig2}.  Here we had over $4000$ data points.  We chose to make predictions at about $100$ data points of which over $90$ we accept.  Again these are the first data points we attempt to make a forecast on.  And again we see some false predictions near the end.  Incorporating a scheme to not make predictions immediately after a false prediction as mentioned in Section \ref{simresult} would leave us with only $1$ false prediction and over $90$ good predictions.

\subsection{Extension to Other Games}

Note that we chose the Minority Game as the game we thought best exhibits the dynamics of the financial time-series we analyzed.  The method for forecasting we describe in this paper can be used with a number of different models (or games).  We require the following of our model and forecast scheme: \\

\noindent 1) We can parameterize the problem into quantities we would like to estimate iteratively. \\

\noindent 2) We have a way to estimate the transition dynamics for the parameters at each iteration and this function will always lie in the class of continuously differentiable functions.  \\

\noindent 3) We have a way to estimate the measurement function which takes the parameter space into the measurement space at each iteration and this function will always lie in the class of continuously differentiable functions. \\

\noindent 4) We have a way to estimate the process noise and measurement noise at each iteration.

\section{Additional Algorithm Tests}

Here we set up some Monte Carlo runs and discuss two additional tests to check that our algorithm is working properly.  The first is a test to check the inner workings of the algorithm, while the second checks the functional results that we are interested in.

\subsection {State Estimate Errors}

As a further validation to the method described in this paper, we would like to check that the state estimate errors \eqref{se2} are tending towards a zero mean process.

We set up the problem exactly as in Section \ref{appsim} and do 400 Monte Carlo runs where our Monte Carlo space is the initial time horizon ($2^{50}$ possibilities) and our initial distribution used for generating the simulation (infinite possibilities) as described in Section \ref{appsim-gsd}.  For each run, we pick the time horizon uniformly at random and we pick the initial distribution by taking a random vector with each element chosen uniformly at random from [0,1] and then normalizing the vector.

\begin{figure}[h!] 
\begin{center}
\psfrag{Monte Carlo Runs}{Monte Carlo Runs}
\psfrag{State Residuals}{State Residuals}
\psfrag{Times}{Times}
\includegraphics[height=4in,width=\textwidth]{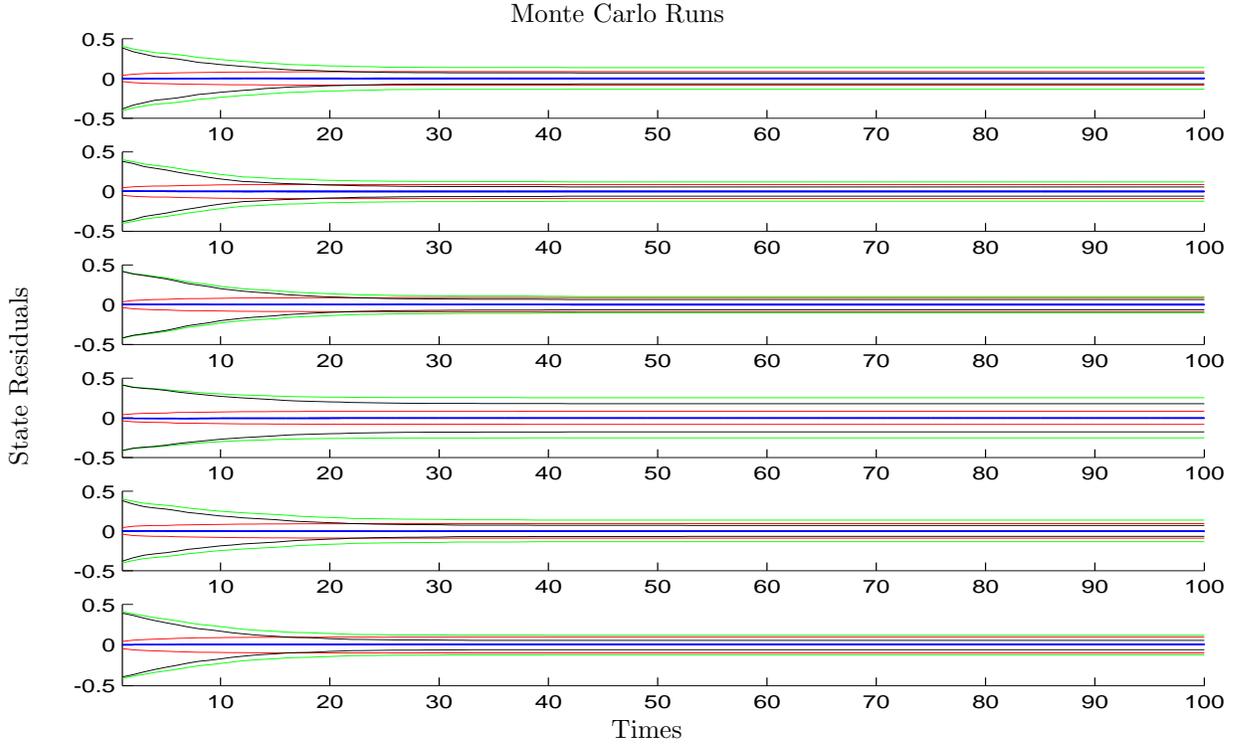}
\end{center}
\caption{In each of the above plots, the blue line indicates the mean innovation process with the error bars indicating the standard error of the mean as given over 400 runs.  The red line indicates the calculated standard deviation based on the sample statistics centered around zero.  The green line indicates the mean standard deviation calculated as the square root of the mean variance over the 400 runs.  The black line indicated the mean standard deviation calculated as the mean of the standard deviations over the 400 runs.}
\label{fig3}
\end{figure}

In order to determine the state estimate errors, we take as the true state the expectation of our true state $\frac{1}{6}1_{6}$ which is also our educated guess for our initial state from Section \ref{ip}.  Notice that the red line in Figure \ref{fig3} actually starts out initially with a very tight covariance.  This is a result of choosing our initial state as the expectation of the true state (but not the true state).

\subsection{Innovations Process}

\begin{figure}[h!] 
\begin{center}
\psfrag{Monte Carlo Runs}{Monte Carlo Runs}
\psfrag{Innovations}{Innovations}
\psfrag{Times}{Times}
\includegraphics[height=4in,width=\textwidth]{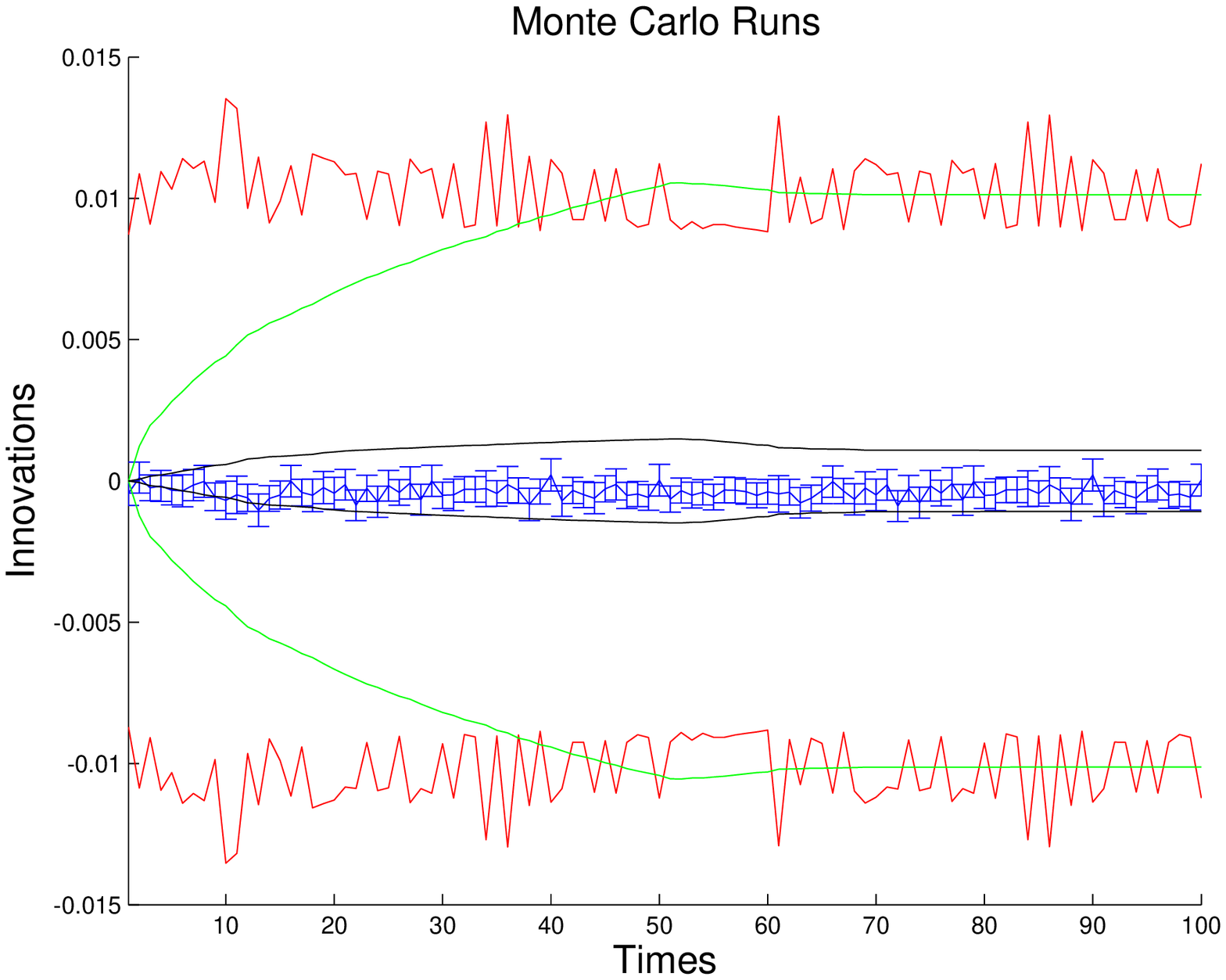}
\end{center}
\caption{In the above plot, the blue line indicates the mean innovation process with the error bars indicating the standard error of the mean as given over 400 runs.  The red line indicates the calculated standard deviation based on the sample statistics centered around zero.  The green line indicates the mean standard deviation calculated as the square root of the mean variance over the 400 runs.  The black line indicated the mean standard deviation calculated as the mean of the standard deviations over the 400 runs.}
\label{fig4}
\end{figure}

The other test that would be important to us is to ensure that our forecast method would give innovations that have mean zero also.  This is a given based on Figure \ref{fig3}.  We show this result in Figure \ref{fig4}.  Notice that the parabolic nature of the green and black lines here is due to the fact that we take our first measurement to have no noise as mentioned in Section \ref{ip}.

\begin{figure}[h!] 
\begin{center}
\psfrag{Monte Carlo Runs (Predictions)}{Monte Carlo Runs (Predictions)}
\psfrag{Innovations}{Innovations}
\psfrag{Times}{Times}
\includegraphics[height=4in,width=\textwidth]{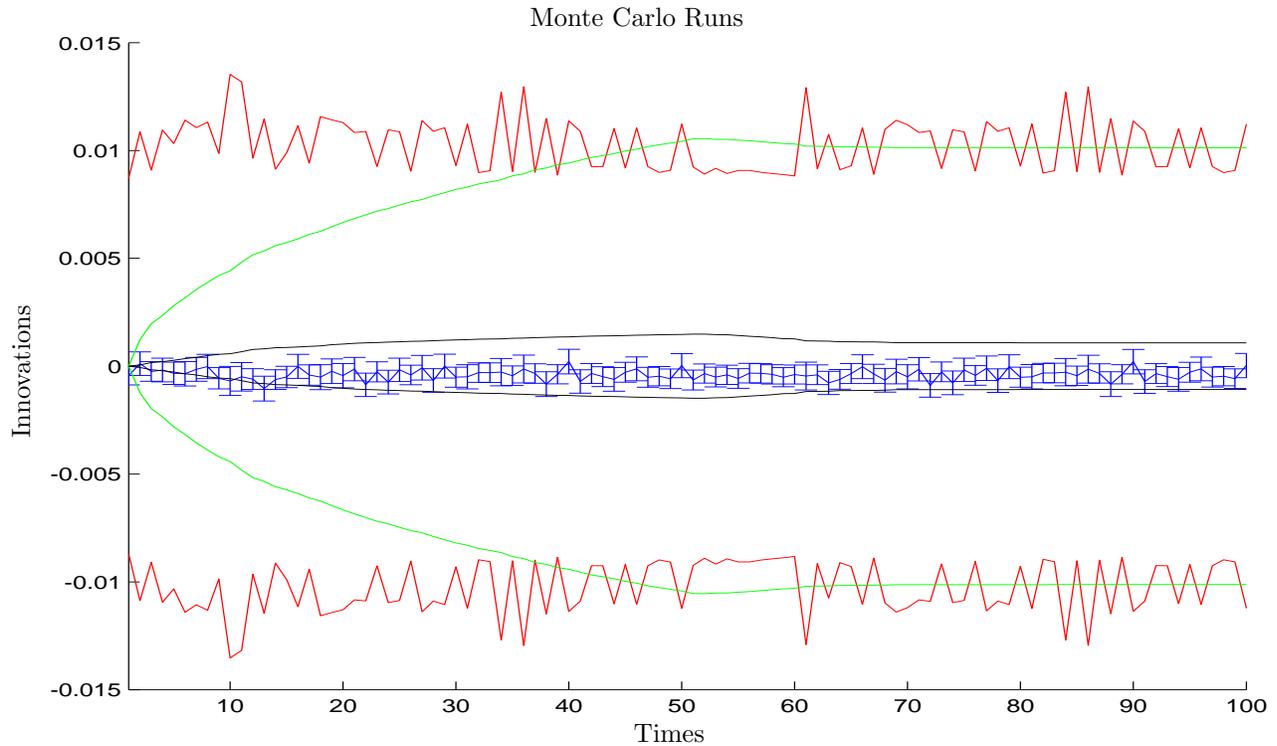}
\end{center}
\caption{In the above plot, the blue line indicates the mean innovation process with the error bars indicating the standard error of the mean as given over 400 runs.  The red line indicates the calculated standard deviation based on the sample statistics centered around zero.  The green line indicates the mean standard deviation calculated as the square root of the mean variance over the 400 runs.  The black line indicated the mean standard deviation calculated as the mean of the standard deviations over the 400 runs.}
\label{fig5}
\end{figure}

Finally, in Figure \ref{fig5}, we show one more plot in which at each time step, we keep only those points which have $S_{k} \leq t_{k} = 10^{-3}$.  In any given time step, at most $3$ points of the $400$ were removed.  These would correspond to difficult to estimate problems (such as the example we chose for Figure \ref{fig1}).  We notice that removing these points tightens our covariances drastically as hoped.

\section{Conclusion}

We have shown a way to forecast time-series by parameterizing an artificial market model such as the Minority Game, using an iterative numerical optimization technique.  In our technique we also describe how to follow an effective forecasting scheme which leads to pockets of predictability.

This paper is meant only to serve as an introduction to such methods of forecasting.

\nocite{*}

\bibliography{JEIC}

\end{document}